\documentclass[12pt]{revtex4}
\usepackage{amsmath,amssymb,bm}
\usepackage[toc,page]{appendix}
\usepackage{graphicx}
\usepackage{url}
\usepackage{hyperref,xcolor}

\DeclareGraphicsExtensions{.pdf,.png,.jpg,.eps}

\begin{document}
\title{Microscopic origin of Boson peak in two-dimensional scalar spring network}
\author{Cunyuan Jiang}
\email{cunyuanjiang@zzu.edu.cn}
\affiliation{School of Physics, Zhengzhou University, Zhengzhou 450001, Henan, China}

\begin{abstract}
We proposed a non-analytic model to explain the microscopic origin of the anomalous vibrational density of states (DOS), the Boson peak (BP), in amorphous solids based on the scalar dynamical matrix of a network with springs and nodes. We argue that disorder can be classified into two factors: fluctuation of spring strength and fluctuation of coordination numbers (the number of springs connected to a node). The results suggest that BP originates from fluctuation of coordination numbers, while the fluctuation of spring strength only contributes to the effect of damping and has very limited effect on low frequency DOS. This work converts complexity into simplicity and provides a direct answer to the puzzle of the microscopic origin of BP in amorphous solids.
\end{abstract}

\maketitle
\section{Introduction}
It has been over half a century since the seminal work reported anomalous low temperature thermodynamical properties of amorphous solids compared with crystalline counterparts\cite{PhysRevB.4.2029}. Yet when we ask the question what is the origin governing the anomalous low temperature and low frequency features of amorphous solids, the answers are still much controversial. The anomalous feature of amorphous solids is usually referred to as the 'Boson peak (BP)' in the literature, characterized by additional vibrational density of states (DOS) beyond Debye's theoretical perspective\cite{debye}. Therefore the additional DOS will contribute to a peak (which is the BP) in the low frequency region if one plots the DOS reduced by Debye's law, \(g(\omega) / g_{Debye}(\omega)\). Various opinions have been developed from different points of view to grasp the nature of disorder and to interpret the mechanism leading to BP. For example, the points of view include mainly the anharmonicity of interatomic potential\cite{PhysRevB.67.094203,PhysRevB.43.5039,PhysRevB.46.2798,PhysRevB.49.9400}, the spatial fluctuation of elasticity\cite{PhysRevLett.81.136,Marruzzo2013}, and the presence of phenomenological dynamical objects\cite{Hu2022,PhysRevLett.133.188302,C2SM26789F,C2SM27533C,Corrado2020,PhysRevLett.108.095501,PhysRevResearch.6.023053}. However, beyond the continuum and dynamics point of view, the microscopic structure at the atomic scale, which should be one of the most fundamental features of amorphous solids, is relatively rarely discussed in the puzzle of BP's origin. 

There are three featured dynamical anomalies of amorphous solids\cite{Hu2022}: BP, fast \(\beta\)-relaxation\cite{Wang2015}, and slow structural relaxation\cite{Zhao2025,doi:10.1073/pnas.1120147109}. The latter two involve the rearrangement of particles, which means interactions between particles can disappear and be created as time goes by. BP does not involve such rearrangement of particles, which is the basis of analysis (allowing the) use of the dynamical matrix (Hessian matrix\cite{Hu2022,ch39-6bhs} or inverse of the correlation matrix\cite{PhysRevLett.133.188302}) in the community. Once the Hessian matrix is obtained, BP and the static atomic structure of the system can be determined directly by statistics of the distribution of the eigenfrequencies of the matrix and by examining the positions of non-zero matrix elements. Therefore, the secrets of BP are fully hidden in the structure of the dynamical matrix. There are two kinds of 'degrees of freedom' for each particle in the dynamical matrix that can be used to break the translational symmetry between particles and hence create 'disorder' in the system. The first is the strength of interaction between this particle and the others, characterized by the value of non-zero off-diagonal elements. The second is the number of particles having interaction with this one, the coordination number\cite{PhysRevLett.98.175502}, characterized by the number of non-zero off-diagonal elements in the dynamical matrix. The values of off-diagonal elements, if they are not constant, can be considered to relate to spatial fluctuation of elasticity\cite{PhysRevLett.81.136,Marruzzo2013,Jiang_2026}, although the crossover from discrete particles to the continuum limit should be treated very carefully. The effect of coordination number on the creation of anomalous vibrational modes is relatively less reported\cite{PhysRevLett.98.175502,PhysRevB.83.184205,PhysRevB.93.094204}.

In this work, we simplify the spatial degrees of freedom of the dynamical matrix and model the amorphous solid as a network of springs and nodes. The simplification will lead to missing the information of in-plane (for two-dimensional situations) dynamics (longitudinal or transverse) but highlights the two factors mentioned above: the strength of springs and coordination numbers. The results show that spatial fluctuation of spring strength alone is not going to induce additional DOS beyond Debye's prediction; however, spatial fluctuation of coordination number is the factor that solely determines anomalous DOS in the low frequency region. We also provide simulation data to confirm the distribution of coordination numbers in the model. And finally, we show that whether the coordination number of a node is larger or smaller than the average level will lead to additional dynamical response (DOS) at new frequencies. Our work clearly suggests that fluctuation of coordination number on different particles serves as the microscopic origin of BP in amorphous solids.

\section{Model construction}
\begin{figure}
    \centering
    \includegraphics[width=\linewidth]{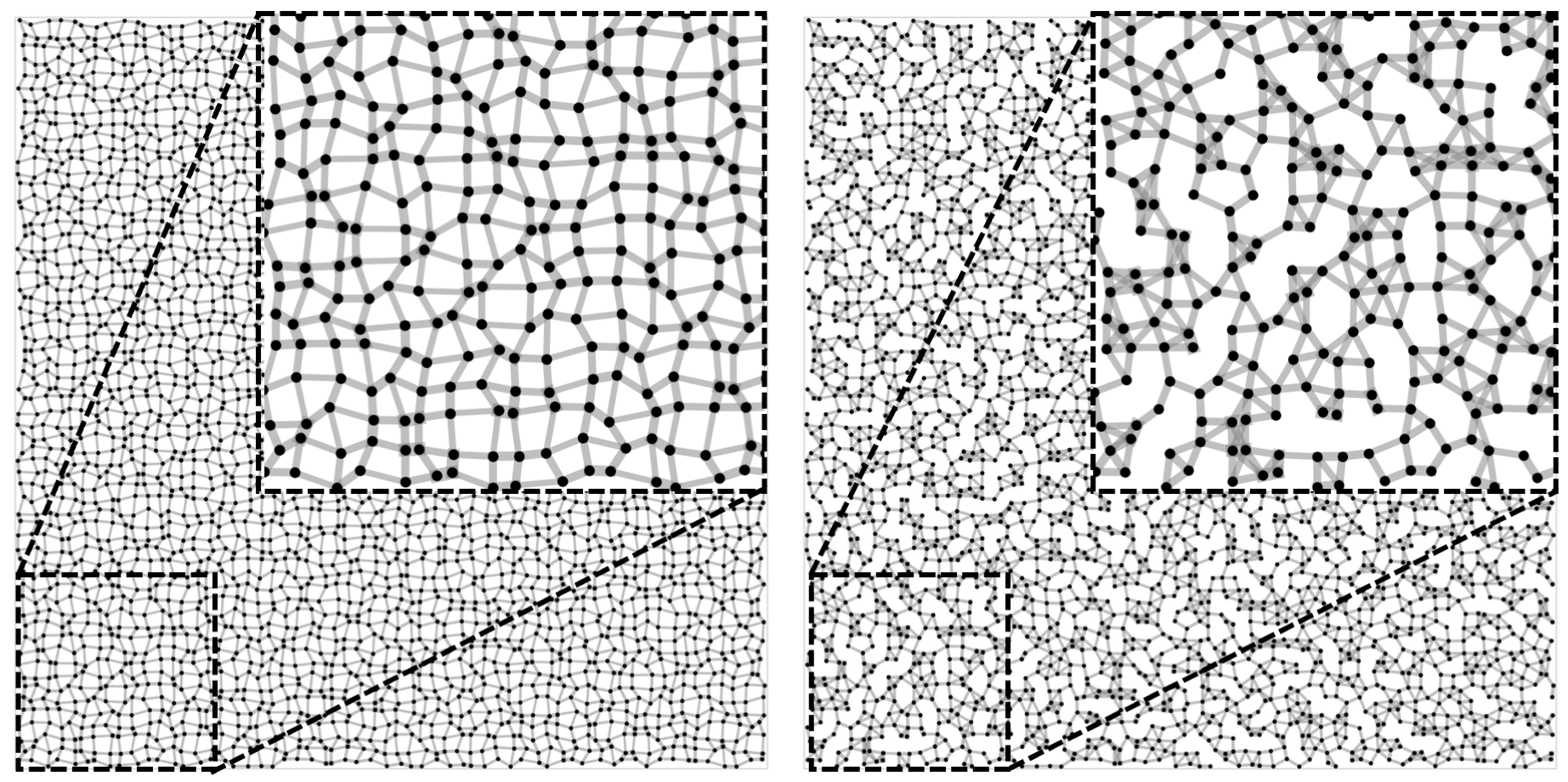}
    \caption{Structures of two-dimensional springs (gray lines) network where each node (black dots) has four springs connected with others (coordination number) (\textbf{left}), and where each node has an undetermined number of springs connected with others according to their distance (\textbf{right}). The positions of nodes are generated based on a square lattice (lattice constant \(a = 1\)) with a fixed distance of deviation \(\delta\) from the lattice site and along a random direction \(\theta \in [0, 2 \pi]\). In the left panel, the springs are connected between the nearest neighbors of the square lattice sites. In the right panel, the springs are connected between nodes with distance smaller than \(a + \delta\). The left bottom corner has been enlarged as shown in the dashed boxes. Periodic boundary conditions are applied in the analysis throughout this work.}
    \label{figstructures}
\end{figure}

A network of springs and nodes is needed to construct a corresponding dynamical matrix \(H\). For a \(\sqrt{N} \times \sqrt{N}\) two-dimensional square lattice with lattice constant \(a\), there are \(N\) nodes in total. The dynamical matrix, if spatial degrees of freedom are simplified, is of size \(N \times N\). The off-diagonal elements are minus the spring strength \(H_{ij(i \neq j)} = -t_{ij}\) between the \(i\)th and \(j\)th nodes. Most off-diagonal elements are zero, except few connected neighbors. The diagonal elements are \(H_{ii} = \sum_{j \neq i} t_{ij}\) to ensure force balance along all rows and columns of the matrix. To approach the situation of amorphous solids, we allow all nodes to deviate from square lattice sites by a distance \(\delta\) (same for all nodes) smaller than the lattice constant \(a\) (which can be normalized to be \(1\)) along an independent random direction for each node \(\theta_i\) as shown in Fig.\ref{figstructures}. The random direction \(\theta_i\) uniformly distributed in \(\in [0, 2 \pi]\). To grasp the spatial fluctuation of spring strength, we assume that the strength of springs is proportional to the inverse of the distance between the two nodes \(r_{ij}\), \(t_{ij} = t_0 / r_{ij}\), with \(t_0 = 1\). Such an interaction intensity is an assumption to generate structure dependent spatial fluctuation. To avoid \(1/0\) unrealistic interaction intensity, the parameter setup must satisfy \(a - 2 \delta >0\). In following numerical analysis, we set \(\delta = 0.3 a\) as the largest \(\delta\) to avoid unrealistic situation. We will discuss two situations to separate the effect of spatial fluctuation of spring strength and that of coordination numbers.

The model simplifies two spatial degrees of freedom \(x\) and \(y\) of a two-dimensional system into one. By doing so, the direction of each spring is simplified, only the information of connectivity and intensity are reserved. Given that BP is neither contributed solely by transverse nor longitudinal vibrations as reported by Ref.\cite{Xi2026}, the simplification of spatial degrees of freedom is helpful to let us focus on the other two factors, connectivity and intensity of springs. 

The first situation, as shown in the left panel of Fig.\ref{figstructures}, is that all nodes are connected by springs as the nearest neighbors of the square lattice. The coordination number is therefore \(4\) for all nodes. The strength of springs is the only variable according to the distance between two nodes. The periodical boundary condition is applied throughout the numerical analysis.

The second situation, as shown in the right panel of Fig.\ref{figstructures}, is that all pairs of nodes with distance shorter than \(a + \delta\) will be connected by a spring. The strength of springs depends on their distance as before. All springs longer than \(a + \delta\) will disappear. In this situation, the coordination number should have a distribution peaked at the average value. The choice of cut off \(a + \delta\) serves for generating peaked distribution of coordination number.

We also assume the mass of nodes is constantly \(1\), and hence the DOS can be obtained directly by statistics of the distribution of eigenfrequencies, the square roots of eigenvalues of the dynamical matrix \(H\).

\section{Results and discussion}
\begin{figure}
    \centering
    \includegraphics[width=\linewidth]{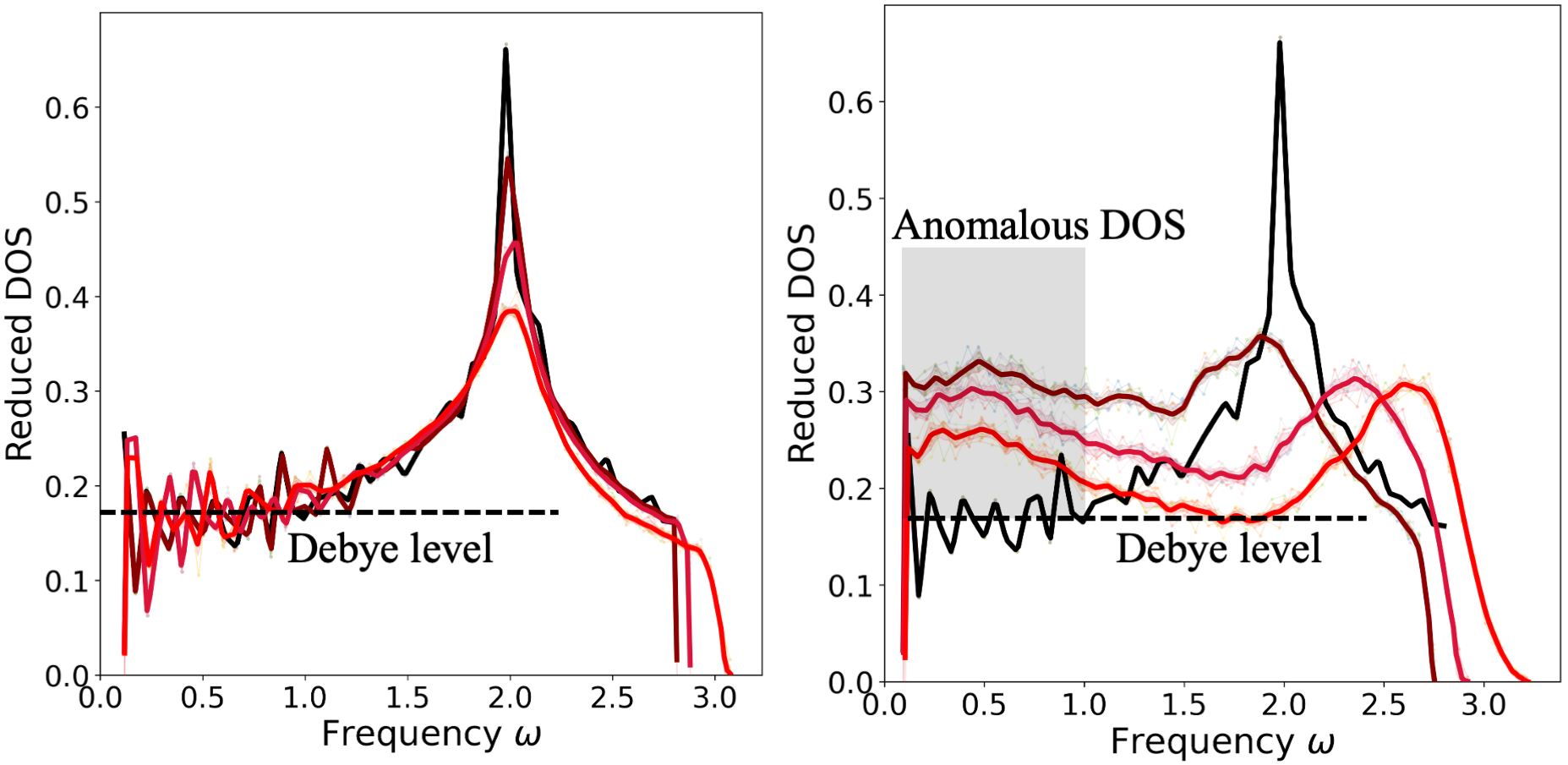}
    \caption{The vibrational density of states (DOS) reduced by Debye's law \(g(\omega) / \omega\) for \(\delta = 0, \, 0.1, \, 0.2, \, 0.3\) from the thick black line to the red line with coordination number of each node fixed to be \(4\) (\textbf{left}) as shown in the left panel of Fig.\ref{figstructures}, and unfixed determined by the distance (\textbf{right}) as shown in the right panel of Fig.\ref{figstructures}. Each parameter of \(\delta\) has been averaged over ten times with different random seeds; the non-averaged results are shown by the thin lines around the thick lines. All curves in the left panel obey Debye's law in the low frequency region, \(\omega < 1\). In the right panel, the anomalous DOS with respect to Debye's law has been highlighted in gray. In the numerical computation, a \(60 \times 60\) grid is used.}
    \label{figdos}
\end{figure}

The DOS reduced by Debye's law are shown in the left and right panels of Fig.\ref{figdos} corresponding to the first and the second situations mentioned above. Four different parameters of deviation from the lattice site \(\delta = 0, \, 0.1, \, 0.2, \, 0.3\) are applied to illustrate the effect of different disorder levels. \(\delta = 0\) is the case of a crystal for reference and free from any disorder. In the left panel of Fig.\ref{figdos}, it can be seen that the effect of spatial fluctuation of spring strength is limited only to the high frequency region to weaken the Van Hove peak (induced by the Van Hove singularity where the wavelength is close to the lattice constant). Debye's prediction for low frequency is always valid. 

In the right panel of Fig.\ref{figdos}, the reduced DOS for the four parameters \(\delta\) as in the left panel of Fig.\ref{figdos} are shown. Here, the coordination numbers are not fixed at \(4\) but are allowed to change. The strength of springs is also allowed to change according to the distance between nodes. The results exhibit many features that should be noticed. The most important feature is the emergence of additional DOS beyond Debye's law in the low frequency region roughly \(\omega \in [0,1]\) (colored in gray). The additional DOS peaks at \(\omega = 0.5\) as the position of BP for all three situations where \(\delta \neq 0\). The second feature is that the Van Hove peak does not disappear under disorder. The coexistence of BP and the Van Hove peak agrees with previous experimental investigations\cite{PhysRevB.98.174207} and supports that BP and the Van Hove peak are different phenomena. 

\begin{figure}
    \centering
    \includegraphics[width=\linewidth]{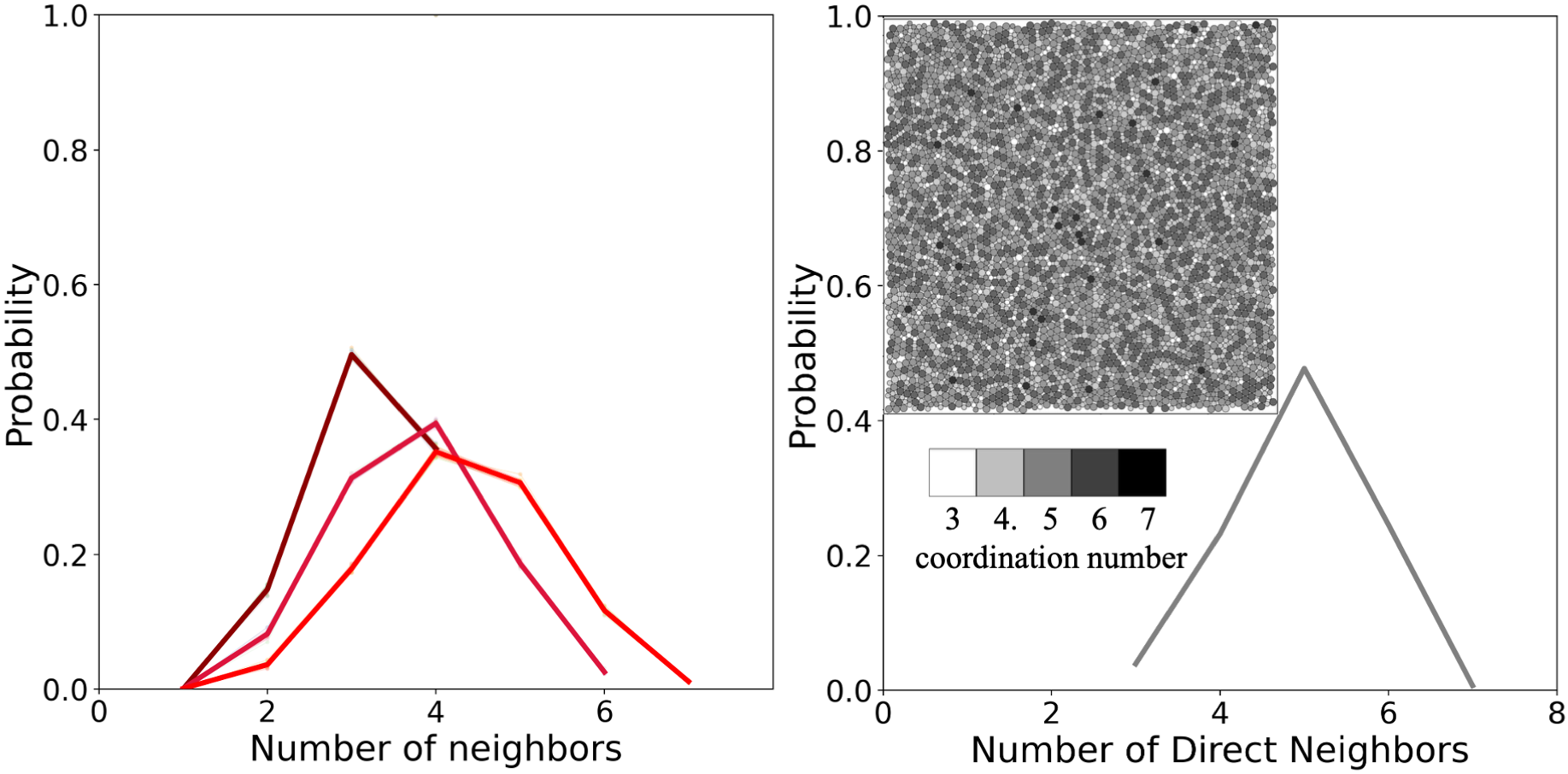}
    \caption{\textbf{Left}, the distribution of coordination numbers (number of neighbors) for \(\delta = 0.1, \, 0.2, \, 0.3\) from black to red corresponding to the situation in the right panel of Fig.\ref{figdos}. The structure is schematically shown in the right panel of Fig.\ref{figstructures}. Each curve has been averaged over ten times with different random seeds. \textbf{Right}, the distribution of coordination number of a simulated two-dimensional amorphous solid (Hessian matrix)\cite{jiang2026bosonpeakphenomenonparticipated}. The inset shows the structure of the simulated amorphous solid. Particles are colored according to their coordination number from white to black from \(3\) to \(7\).}
    \label{figdistribution}
\end{figure}

Properties of DOS and BP are fully determined by the dynamical matrix. For a dynamical matrix, once we know two groups of information, how many non-zero off-diagonal elements there are for each row and column (coordination numbers) and whether they are identical or not (fluctuation of spring strength), the dynamical matrix can be determined. The results of Fig.\ref{figdos} demonstrate that fluctuation of coordination numbers is the dominating factor that gives rise to the BP while fluctuation of spring strength only induces an obvious damping effect in the high frequency region. To compare our model construction with the real situation in amorphous solids, the distribution of coordination numbers of each node should be known. The left panel of Fig.\ref{figdistribution} shows such distributions for \(\delta = 0.1, \, 0.2, \, 0.3\) corresponding to the three situations in Fig.\ref{figdos}. The distribution peaks at around \(4\) due to the number of nearest neighbors of the square lattice. In the right panel of Fig.\ref{figdistribution}, the distribution from a simulated two-dimensional amorphous solid is shown\cite{jiang2026bosonpeakphenomenonparticipated}. It is also a peaked distribution as in our model. Therefore, the model is able to capture the fluctuation of coordination numbers of amorphous solids. 

\begin{figure}
    \centering
    \includegraphics[width=\linewidth]{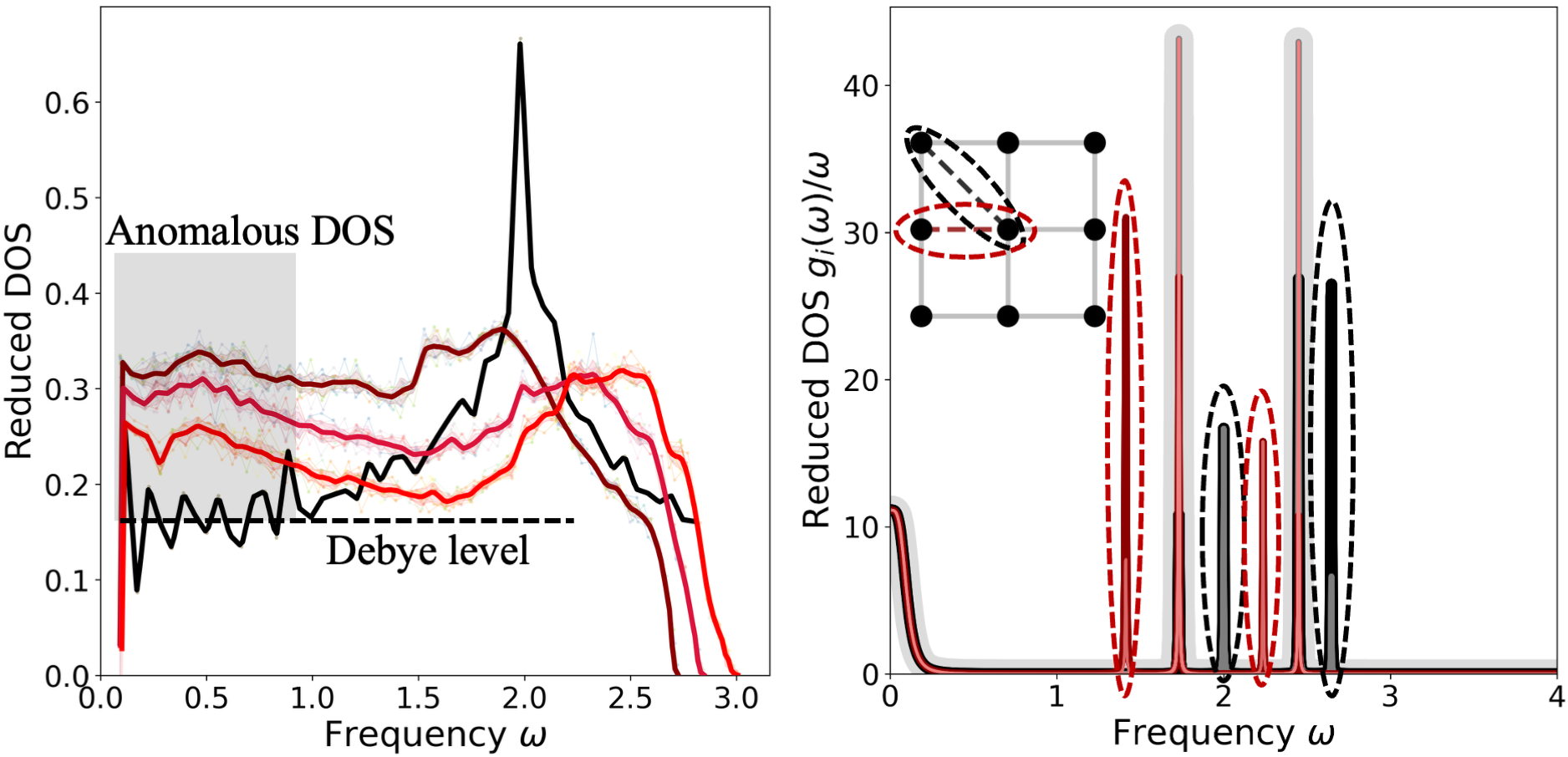}
    \caption{\textbf{Left}, the reduced DOS \(g(\omega) / \omega\) as shown in the right panel of Fig.\ref{figdos}, but the strength of all springs is fixed to be \(1\). \textbf{Right}, the particle-level reduced DOS \(g_i (\omega) / \omega = -\text{Im}[G(\omega)_{ii}]\) of a grid with nine nodes shown in the inset. There are three situations. The first situation is nine nodes connected as a square lattice. The result of this situation is shown by the thick transparent gray lines. The second situation has an additional spring between the center and the top left nodes as shown in the inset. The local DOS of the center and the top left nodes are colored black and the rest are colored thinner gray under this situation. The third situation has a spring missing between the center and the left nodes as shown in the inset. Results are colored darker red (for the two nodes) and red (for the rest) under this situation. The situation of the square lattice has only two dynamical response peaks, poles (apart from the one at zero frequency), while both of the other situations with an additional or missing spring have two additional poles lying in the dashed circles.}
    \label{figtoy}
\end{figure}

In the right panel of Fig.\ref{figdos}, the fluctuation of spring strength and fluctuation of coordination numbers appear together. One may be curious what happens if the strength of springs is fixed and only fluctuation of coordination numbers appears. This test can be easily achieved in our model by requiring the non-zero off-diagonal elements of the dynamical matrix to be a constant, \(H_{ij(i\neq j)} = - t_0\), and then the diagonal elements are directly the coordination numbers of the corresponding nodes, \(H_{ii} = \sum_{j \neq i} t_0\) with \(t_0 = 1\). The left panel of Fig.\ref{figtoy} shows the DOS of this setup with the same random seeds and parameters of \(\delta\) as in the right panel of Fig.\ref{figdos}. By comparing the left panel of Fig.\ref{figtoy} and the right panel of Fig.\ref{figdos}, it can be concluded that the fluctuation of spring strength affects the emergence of BP very little. This test, to exclude fluctuation of spring strength, strengthens the previous conclusion that fluctuation of coordination number should be the microscopic origin of BP. 

To interpret the mechanism of how fluctuation of coordination numbers can induce additional DOS, let us consider a toy model with only nine nodes as shown in the inset of the right panel of Fig.\ref{figtoy}. The right panel of Fig.\ref{figtoy} shows the on-node dynamical response functions of each node for three situations: the springs are connected as a square lattice (gray), there is an additional spring between the center and the top left nodes (black), and the spring between the center and the left nodes is missing (red). Here the on-node dynamical response functions are the imaginary parts of the diagonal matrix elements \(-\text{Im}[G(\omega)_{ii}]\) of the Green's function\cite{jiang2026bosonpeakphenomenonparticipated},
\begin{equation}
    G(\omega) = [(\omega^2 + i \epsilon)I - H]^{-1} ,
\end{equation}
with \(H\) the dynamical matrix, \(\epsilon =  0^+\) a small number, and \(I\) the unit matrix. The imaginary parts of the diagonal elements give the on-node DOS, and they will be non-zero (have peaks) at the poles of the Green's function. In the right panel of Fig.\ref{figtoy}, it can be seen that there are only two peaks for the situation of the square lattice with coordination number fixed to \(4\) (the thick gray lines). For the situation with an additional spring (the thin black and gray lines), two nodes have coordination number \(5\). The dynamical response functions have two peaks overlapping with the situation of no additional springs (gray) and have two additional peaks at other frequencies as highlighted by black dashed circles. The third situation has a spring missing between the center and the left nodes, so the coordination numbers become \(3\) for the two nodes. The results are similar to adding a spring in that two peaks overlap with the crystal situation and two new peaks are created as highlighted by red dashed circles. Additional peaks indicate the presence of additional poles of the Green's function and dynamical modes. Therefore the toy model demonstrates that fluctuation of coordination numbers away from the average number, \(5\) and \(3\) from \(4\), is able to induce additional modes at new frequencies that do not exist before. Meanwhile, the fluctuation of coordination number does not destroy the dynamical modes in the crystal since the two old peaks always survive.

In the right panel of Fig.\ref{figtoy}, attention should be paid to a detail that the additional dynamical peaks are not only contributed by those nodes with fluctuated coordination number (the center node and the other nodes with special springs). The nodes with coordination number \(4\) that do not have additional or missing springs also contribute to the additional peaks as shown by the thicker gray and lighter red lines in Fig.\ref{figtoy}. This result indicates that the additional modes, and hence BP, are extended rather than localized and are participated in by a large ratio of particles\cite{10.1063/5.0147889,Wang2019,PhysRevB.105.014204}. 

The toy model in the right panel of Fig.\ref{figtoy} demonstrates that fluctuation of coordination number is able to induce additional modes, but it cannot yet predict the frequency where the additional modes appear concentrated to form a BP, due to the small number of nodes. By increasing the nodes to the order of a few thousand, the situation becomes that shown in the left panel of Fig.\ref{figtoy} or in the right panel of Fig.\ref{figdos}, where BP is more obvious but the clarity of their creation is lost. Beyond our simplification, the direction of springs on each node may also contribute to BP through other mechanisms such as local inversion symmetry breaking\cite{PhysRevB.83.184205,PhysRevB.93.094204}, or internal stresses\cite{p95w-9wjc}. The mechanisms are not applicable to the current model scalar network due to the miss of direction of springs in here. Therefore, from our perspective, a more analytical theory that explains BP as originating from fluctuation of coordination number of scalar springs network in the low frequency region is still being looked forward to. In the left panel of Fig.\ref{figtoy}, the region of BP appearing in the region \(0 < \omega < 1\) and peaking at \(\omega = 0.5\) is robust against changes of structure, changes of the distribution of coordination numbers, and even the weak fluctuation of spring strength from the average value as shown in the right panel of Fig.\ref{figdos}. Such robustness provides an opportunity to model the region of BP by a few order parameters in the future.  

\section{Conclusion}
Mathematically, BP and the total DOS of amorphous solids are fully determined by the two features of the dynamical matrix: the coordination numbers of each node and the strength of each spring. We simplify the spatial degrees of freedom in the dynamical matrix to model the network of nodes and springs and find that BP appears only when the coordination numbers of nodes do not remain constant for different nodes, which is fluctuation of coordination numbers. The fluctuation of spring strength is not able to affect the low frequency DOS to shape BP beyond Debye's law but contributes a damping effect that weakens the Van Hove peak at higher frequencies. We then compare the distribution of coordination numbers of our model with the simulation data to validate their peaked distribution and hence the authenticity of our model networks. At the end, we show a toy model to demonstrate that deviation of coordination number from the average value is able to create additional dynamical modes in the system. Our work simplifies the dynamics of amorphous solids (Hessian matrix) into two factors, coordination numbers and strength of springs, and identifies the coordination numbers as the microscopic origin of the low frequency anomalous DOS of amorphous solids. The results provide a clear answer to the puzzle of the microscopic origin of BP.

\noindent \textit{Author Contribution } -- The first author designs and performs the research.

\noindent \textit{Data Availability} -- All data are available by contacting the corresponding author.

\noindent \textit{Competing Interests} -- The author declares that there are no competing interests.

\noindent \textit{Funding} -- Not applicable.

\noindent \textit{Acknowledgments} -- 


\end{document}